\documentclass[twocolumn,trackchanges,twocolappendix]{aastex701}

\newcommand\gaia{{\em Gaia\ }}
\newcommand\rguide{$R_{Guide}$~}
\newcommand\rgc{$R_{GC}~$}

\usepackage{lettrine}
\input Zallman.fd

\usepackage{amsmath}
\usepackage{moresize}
\usepackage{soul}
\usepackage{gensymb}
\usepackage[OT1]{fontenc}

\graphicspath{{./}{figures/}}

\begin{document}

\title{The Open Cluster Chemical Abundances and Mapping Survey XI.\\ First Gradients from SDSS/MWM BOSS Determined Clusters}

\correspondingauthor{Jonah M.~Otto}

\author[0000-0003-2602-4302]{Jonah M.~Otto}
\affiliation{Department of Physics and Astronomy, Texas Christian University, TCU Box 298840 
Fort Worth, TX 76129, USA }
\email[show]{j.otto@tcu.edu}

\author[0000-0002-0740-8346]{Peter M.~Frinchaboy}
\affiliation{Department of Physics and Astronomy, Texas Christian University, TCU Box 298840 
Fort Worth, TX 76129, USA }
\email{p.frinchaboy@tcu.edu}

\author[0009-0000-4049-5851]{John Donor}
\affiliation{Department of Physics and Astronomy, Texas Christian University, TCU Box 298840 
Fort Worth, TX 76129, USA }
\email{j.donor@tcu.edu}

\author[0000-0003-3410-5794]{Ilija Medan}
\affiliation{Department of Physics and Astronomy,
	Vanderbilt University,
	Nashville, TN 37235, USA}
\email{ilija.medan@vanderbilt.edu}

\author[0000-0001-9738-4829]{Natalie R.~Myers}
\affiliation{William H.\ Miller III Department of Physics \& Astronomy,
Johns Hopkins University, 3400 N Charles St, Baltimore, MD 21218, USA}
\email{nrmyers.astro@gmail.com}

\author[]{Audrey Hauck}
\affiliation{Department of Physics and Astronomy, Texas Christian University, TCU Box 298840 
Fort Worth, TX 76129, USA }
\email{a.hauck@tcu.edu}

\author[0000-0001-6476-0576]{Katia Cunha}
\affiliation{Steward Observatory, University of Arizona, 933 North Cherry Avenue, Tucson, AZ 85721-0065, USA}
\affiliation{Observatório Nacional/MCTIC, R. Gen. José Cristino, 77,  20921-400, Rio de Janeiro, Brazil}
\email{kcunha@arizona.edu}

\author[0000-0001-6761-9359]{Gail Zasowski}
\affiliation{Department of Physics and Astronomy, University of Utah, 270 S. 1400 E. \#E2108, Salt Lake City, UT 84112, USA}
\email{u0948422@gcloud.utah.edu}

\author[0000-0002-1715-1257]{Madeleine McKenzie}
\affiliation{Observatories of the Carnegie Institution for Science, 813 Santa Barbara St., Pasadena, CA 91101}
\email{mmckenzie@carnegiescience.edu}

\author[0000-0003-2868-8276]{Jingkun Zhao}
\affiliation{National Astronomical Observatories, Chinese Academy of Sciences, Beijing 100101, People's Republic of China}
\email{zjk@nao.cas.cn}

\author[0000-0002-6972-6411]{Jos\'e Eduardo M\'endez-Delgado}
\affiliation{Instituto de Astronom\'{i}a, Universidad Nacional Aut\'{o}noma de M\'{e}xico, A.P. 70-264, 04510, Ciudad de M\'{e}xico, M\'{e}xico}
\email{jmendez@astro.unam.mx}

\author[0009-0005-0182-7186]{Amaya Sinha}
\affiliation{Department of Physics and Astronomy, University of Utah, 270 S. 1400 E. \#E2108, Salt Lake City, UT 84112, USA}
\email{u1363702@utah.edu}

\author[0000-0003-0179-9662]{Zachary Way}
\affiliation{Department of Physics and Astronomy, Georgia State University, Atlanta, GA 30302, USA}
\email{zway1@gsu.edu}


\author[0000-0002-3601-133X]{Dmitry Bizyaev}
\affiliation{Apache Point Observatory and New Mexico State University, P.O. Box 59, Sunspot, NM, 88349-0059, USA}
\affiliation{Sternberg Astronomical Institute, Moscow State University, Moscow, 119234, Russia}
\email{dmbiz@apo.nmsu.edu}

\author[0000-0003-0174-0564]{Andrew R.\ Casey}
\affiliation{Center for Computational Astrophysics, Flatiron Institute, 162~Fifth~Ave., New~York, NY~10010,~USA}
\affiliation{School of Physics and Astronomy, Monash University VIC 3800, Australia}
\email{andrew.casey@monash.edu}

\author[0000-0003-3526-5052]{Jos\'e G. Fern\'andez-Trincado}
\affiliation{Centro de investigaci\'on en Astronom\'ia, Facultad de Ingenier\'ia, Ciencia y Tecnolog\'ia, Universidad Bernardo O’Higgins, Av. Viel 1497, Santiago, 8370993, Chile}
\email{jose.fernandez@ubo.cl}

\begin{abstract}

The Milky Way Mapper program in the fifth generation of the Sloan Digital Sky Survey (SDSS-V/MWM) has observed millions of stars, thousands of them in open clusters.
The Open Cluster Chemical Abundances and Mapping (OCCAM) survey continues to create comprehensive datasets of open clusters and their members in order to constrain Galactic parameters. 
This eleventh contribution from the OCCAM survey is the first to use stellar parameters from stars observed with the optical Baryon Oscillation Spectroscopic Survey (BOSS) spectrograph to determine cluster membership. 
We use data from SDSS-V/MWM's 20th Data Release (DR20) and curate a sample of 1883 stars in 111 open clusters, including 95 not in previous OCCAM samples based on infrared data from the Apache Point Observatory Galactic Evolution Experiment (APOGEE) spectrograph. 
The sample includes 16 clusters with stars observed using both the BOSS and APOGEE spectrographs, and we find consistent agreement in measurements of both [Fe/H] and [$\alpha/M$].  
The BOSS sample includes the majority of the clusters at young ages (Age $< 150$ Myr) that complement the APOGEE sample of primarily older clusters. 
We use the combined BOSS+APOGEE OCCAM sample to constrain the radial metallicity gradient with respect to \rguide ($-0.079 \pm 0.005 \text{ dex kpc}^{-1}$) and \rgc ($-0.082 \pm 0.006 \text{ dex kpc}^{-1}$), which agree well with results from previous OCCAM papers using only APOGEE data.
Finally, the inclusion of the primarily young BOSS clusters has not changed that the OCCAM open cluster sample indicates no significant evolution of this gradient in different mono-age populations.

\end{abstract}




\section{Introduction} 
\label{sec:intro}
\lettrine[]{O}{pen clusters} are key stellar populations for Galactic archaeology, providing good spatial coverage of the Galactic disk at a wide range of ages and metallicities \citep[][and references therein]{friel_95, ladalada}. 
Open clusters are coeval, forming from homogeneous gas clouds at the same time, with member stars having the same age \citep[][]{ladalada}. 
They have also been shown to have homogeneous chemical abundances in both observational studies \citep[e.g.,][]{bovy_2016, sinha_24} and explorations of open clusters in simulated FIRE galaxies \citep[e.g.,][]{bhattarai_24}.  
Clustering algorithms have increasingly been used to determine cluster membership. 
Using data from the ESA's \gaia mission \citep[][]{gaia_tot} as well as spectroscopic datasets. 
The first approach has resulted in several large catalogs of likely cluster members \citep[e.g.,][]{cg_18, cg20, HUNT_2023, Hunt2024} greatly increasing the number of known open clusters in the Milky Way, while the latter approach verifies the veracity of the larger studies \citep[e.g.,][]{guerco25}. 
Large spectroscopic surveys such as the Apache Point Observatory Galactic Evolution Experiment \citep[APOGEE][]{apogee}, GALactic Archaeology with HERMES \citep[GALAH][]{galah}, Large sky Area Multi-Object fiber Spectroscopic Telescope \citep[LAMOST][]{lamost}, and Gaia-ESO \citep[][]{gaia_eso} surveys have observed thousands of these stars, allowing for spectroscopic confirmation of membership. 

Spectroscopic catalogs of open cluster members and their bulk parameters can be used to calibrate large surveys \citep[e.g.,][]{meszaros_2025, kos25} as well as constrain Galactic properties such as the radial metallicity gradient \citep[e.g.,][S. Bijavara Sheshashayana et al., {\it submitted}]{janes_79, frinchaboy_13, cunha_16_grads, magrini_2017, occam_p4, spina_21, myers_2022, occasoV, guerco25, yang2025, otto_26}. 
Other tracers have also been used to constrain the radial metallicity gradient, such as classical cepheids \citep[e.g.,][]{korotin_2011, trentin_2023, nunnari_26} and HII regions \citep[e.g.,][]{afflerbach_1996, martin_2017, mendezdelgado_22}, providing comparisons to gradients derived from the youngest open clusters. 

The range of open cluster ages allows for investigations into how the radial metallicity gradient has changed over time, with recent studies e.g., \citet{magrini_23, occasoV, yang2025, otto_26} indicating some variation in the gradients of young and old open clusters. 
It has also been shown that stars, and thus clusters, undergo some amount of radial migration throughout their lifetimes \citep[][]{Sellwood2002}.
Studies of open clusters in simulated FIRE galaxies have also shown that old open clusters can survive interactions that significantly alter their orbital parameters \citep[][]{bob_25}. 
However, it is currently unclear to what degree gradients derived from the oldest open clusters are influenced by radial migration, but \citet{vv_radmig_23} have shown it is a factor that should be considered when deriving Galactic properties. 

This work provides a supplemental optical data OCCAM catalog of bulk cluster parameters and membership probabilities for 1883 stars in 111 open clusters. We outline the data products used in this work in Section \ref{sec:data}. An overview of our methodology is in Section \ref{sec:methods}. In Section \ref{sec:boss_sample} we describe the BOSS OCCAM sample and available Value Added Catalog (VAC). Results are presented in Section \ref{results} and discussed in Section \ref{sec:discussion}. We wrap up with conclusions in Section \ref{sec:conclusions}.

\section{Data} \label{sec:data}

The fifth Sloan Digital Sky Survey \citep[SDSS-V;][]{sdssv_2026} has observed millions of stars as part of the Milky Way Mapper (MWM; J. Johnson et al., {\em in prep}) project, providing spectroscopic information to supplement the kinematic and positional data from the ESA's \gaia mission \citep{gaia_tot}. 
We combine the third \gaia data release \citep[][]{gaiadr3} with the latest MWM data release (DR20; SDSS Collaboration et al. {\em submitted}) to create a uniform sample of high-probability open cluster members.

\subsection{MWM/BOSS DR20 and the BOSS-CLAM}

The SDSS-V/MWM data that is new in DR20 consists of low resolution (R$\approx$1800) optical spectra observed with two identical BOSS \citep[][]{boss_spec} spectrographs.
No new infrared data is released in DR20, but will be included in subsequent SDSS-V data releases.
The Sloan Foundation Telescope at the Apache Point Observatory \citep[][]{sloan_telescope} hosts one of the spectrographs, while the other is mounted on the Ir\`en\`ee du Pont telescope at the Las Campanas observatory in Chile \citep[][]{du_pont} providing dual hemisphere coverage of the sky. 
In 2021 SDSS-V implemented the Focal Plane System (FPS) \citep{FPS, fps_medan} to replace the plug-plate fiber positioner utilized by previous SDSS surveys. 
The FPS robotically places the fibers that feed the spectrographs, allowing for faster turnaround between field designs. 
Targeting information for MWM was first detailed in \citet{dr18} and has been updated for DR20, these changes are discussed in the DR20 paper (SDSS Collaboration et al. {\em submitted}).
All data from the two BOSS spectrographs were reduced with the {\tt IDLspec2D} v6\_2\_1 pipeline \citep[][S. Morrison et al., {\it submitted}]{bolton_idlspec, dawson_idlspec}. 

We use the stellar parameters, ($T_{eff}$, $log(g)$, $[Fe/H]$, $[\alpha/M]$), from the BOSS-CLAM\footnote{We use the BOSS-CLAM v1.0.0, which was released as a Value Added Catalog in SDSS-V/DR20.} (I. Medan et al. {\it submitted}), a specialized version of the Constrained Linear Absorption Model \citep[CLAM,][]{casey2026}, in this work. BOSS-CLAM is a forward model, which maps stellar labels to Non-negative Matrix Factorization (NMF) basis vector weights via a polynomial mapping jointly optimized with the spectral decomposition. This provides a more flexible framework for working with the lower-resolution BOSS data.
The BOSS-CLAM is trained on high-quality labels from the 19th SDSS-V Data Release ASPCAP catalog \citep[DR19,][]{aspcap, dr19, meszaros_2025}, the {\tt BOSS-MINESweeper} VAC \citep[][]{minesweeper}, wide binaries from \citet{elbadry2021}, and the hot star validation catalog from A. Tkachenko et al. ({\it in prep}), to broadly cover the HR diagram. BOSS-CLAM was specifically vetted on a number of open and globular clusters, where it was shown that the abundances were consistent across the Kiel diagram for a wide range of metallicities.
For a complete discussion on BOSS-CLAM, including model architecture, training, and parameter validation please see I. Medan et al., {\it submitted}.

\begin{deluxetable}{lrrrrrr}
\tablecaption{Parameters for Clusters in both the DR19 APOGEE OCCAM Sample and the DR20 BOSS OCCAM Sample}
\label{tab:overlap}
\tabletypesize{\scriptsize}
\tablehead{
   \colhead{Cluster} & 
   \colhead{[Fe/H]} &
   \colhead{[$\alpha$/M]} &
   \colhead{Num} &
   \colhead{[Fe/H]} &
   \colhead{[$\alpha$/M]} &
   \colhead{Num}\\[-3ex]
   \colhead{Name} &
   \colhead{APOGEE (dex)}  &
   \colhead{APOGEE (dex)}  &
   \colhead{Stars}     &
   \colhead{BOSS-CLAM (dex)}  &
   \colhead{BOSS-CLAM (dex)}  &
   \colhead{Stars}
}

\startdata
Alessi 2             & +0.047 & -0.075 & 2 & -0.01 & +0.01 & 9 \\
Alessi 21            & +0.001 & -0.018 & 1 & -0.07 & +0.05 & 15 \\
IC 4756              & +0.047 & -0.071 & 1 & -0.12 & +0.05 & 19 \\
King 1               & -0.042 & -0.043 & 3 & -0.13 & -0.04 & 3 \\
King 6               & +0.043 & -0.050 & 1 & -0.17 & +0.06 & 9 \\
Melotte 71           & -0.137 & -0.036 & 6 &  -0.20 & +0.03 & 7 \\
NGC 188              & +0.080 & -0.013 & 42 & -0.04 & -0.03 & 3 \\
NGC 1901             & -0.144 & -0.041 & 1 & -0.11 & +0.04 & 6 \\
NGC 2168             & -0.028 & -0.046 & 57 & -0.08 & +0.01 & 235 \\
NGC 2423             & -0.011 & -0.057 & 4 & -0.04 &  -0.00 & 4 \\
NGC 2632             & +0.163 & -0.052 & 61 & +0.14 & -0.04 & 25 \\
NGC 2682             & +0.009 & -0.012 & 190 & -0.03 &  +0.00 & 190 \\
NGC 4337             & +0.209 & -0.034 & 4 & +0.21 & -0.02 & 16 \\
NGC 6791             & +0.305 & -0.001 & 59 & +0.29 & -0.02 & 13 \\
NGC 6819             & +0.039 & -0.038 & 40 & +0.06 & -0.04 & 8 \\
Stock 2              & -0.049 & -0.017 & 2 & -0.06 & +0.01 & 14 \\
\enddata

\end{deluxetable}

\subsection{{\it Gaia} Based Cluster Membership and Parameters}

We use the star cluster membership catalog from \citet{Hunt2024} which is updated from \citet{HUNT_2023} as the basis for our analysis. 
\citet{Hunt2024} use the Hierarchical Density-Based Spatial Clustering of Applications with Noise algorithm and \gaia 5-D astrometry (RA, Dec, PMRA, PMDec, and Parallax) to recover clusters and determine membership probabilities for individual stars. 
Other cluster parameters, log(age) and distance, are taken from \citet{cavallo24} who reanalyzed the \citet{Hunt2024} cluster sample.

\subsection{Overlap with APOGEE}

\begin{figure*}[htb]
    \centering
    \epsscale{1.0}
    \includegraphics[width=.95\textwidth]{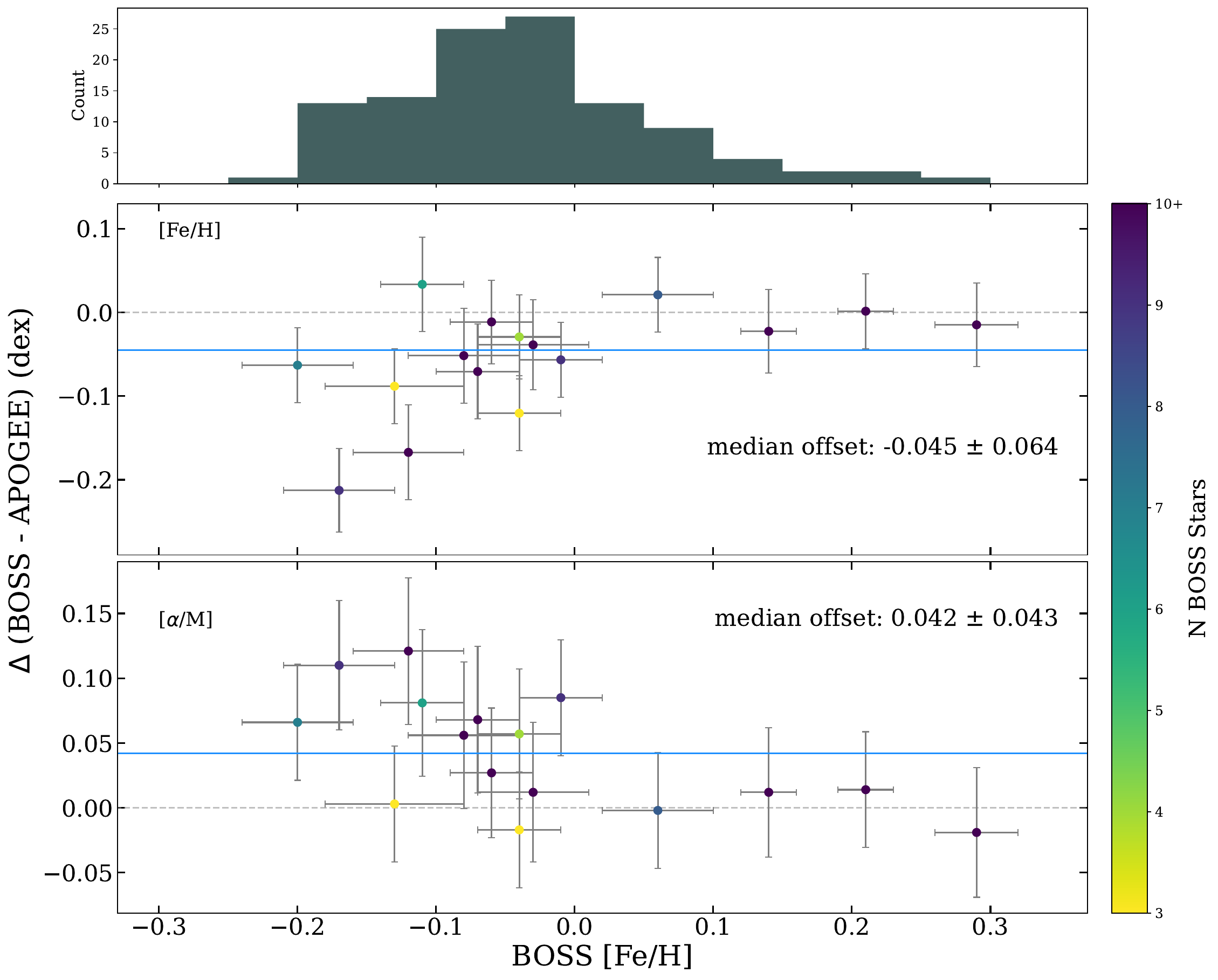}
    \caption{\small The DR20 BOSS minus DR19 APOGEE delta plot for [Fe/H] (top scatter plot) and ([$\alpha$/M] (bottom scatter plot) using the 16 clusters that are in both catalogs. 
    Points are colored by the number of stars in the cluster in the DR20 OCCAM catalog.
    The grey dashed line shows the zero point and the solid blue line shows the median offset. 
    The [Fe/H] distribution of the entire DR20 OCCAM sample is shown above the scatter plots.}
    \label{fig:comp}
\end{figure*}

The 16 clusters in common between the DR20 BOSS OCCAM sample and DR19 APOGEE OCCAM sample are listed in Table \ref{tab:overlap} along with the [Fe/H] and [$\alpha$/M] abundances and number of stars in each cluster. 
We show the graphical form of this table in Figure \ref{fig:comp}, where we find median offsets of $-0.045 \pm 0.064$ dex in [Fe/H] and $+0.042 \pm 0.043$ in [$\alpha$/M]. 
Both abundances have offsets less than the 1-$\sigma$ uncertainty in the measurement, and are smaller than the 1-$\sigma$ uncertainty when comparing BOSS-CLAM [Fe/H] values to ASPCAP [Fe/H] values from the DR19, $0.13$ dex. 
In particular, the four clusters (NGC 2632, NGC 4337, NGC 6791, and NGC 6819) with super-solar [Fe/H] abundances show very good agreement in both [Fe/H] and [$\alpha$/M]. 
Clusters at lower metallicities tend to agree worse than their more metal-rich counterparts, indicating a potential bias. 
In particular we note that clusters deficient in [Fe/H] are enhanced in [$\alpha$/M] as the BOSS-CLAM fits the whole spectra at once causing this anti-correlation to occur, particularly at lower metallicities.
In an effort to correct this effect, we fit a quadratic function to the [Fe/H] vs $\Delta$[Fe/H] points and used the resulting equation to apply a corrective term to all the DR20 BOSS cluster [Fe/H] abundances. 
We show the results of this correction and the resulting gradients in Appendix \ref{app:cal}. 
We find that the corrective measures have little to no effect on the inferred metallicity gradients for both the whole sample and the mono-age populations, and use the abundances as is for the remainder of this work.
The histogram in Figure \ref{fig:comp} shows the [Fe/H] distribution for all clusters in the BOSS OCCAM sample. 
The majority of the clusters have [Fe/H] abundances greater than $-0.1$ dex ($>70\%$), which is where the issue becomes particularly pronounced. 
For a star-by-star comparison between BOSS-CLAM parameters and ASPCAP, as well as other large spectroscopic surveys and a full discussion of this degeneracy, see I. Medan et al., {\it submitted}. 
Overall, there are 9 clusters that are within 1-$\sigma$ of zero offset in both [Fe/H] and [$\alpha$/M] showing there is reasonably good agreement between the lower resolution BOSS abundances and the APOGEE abundances, and that adding the 95 new clusters into our sample will enhance the Galactic gradient discussion. 

\section{Methods} \label{sec:methods}

We determine cluster membership by first selecting all stars in the BOSS-CLAM catalog within $3 \times R_{tot}$\footnote{The total radius of the cluster including tidal tails from \citet{Hunt2024}.} for each cluster.  
We then use the \citet{Hunt2024} member catalog to produce a sample of likely cluster members with BOSS-CLAM data.
While these likely members are determined using 5-D \gaia astrometry, for conciseness we refer to them as proper motion members, and adopt the \citet{Hunt2024} probability as the proper motion probability.
We enhance the \gaia determined proper motion members by adding RV and [Fe/H] probabilities using the Gaussian kernel smoothing routine implemented by \citet{donor_18}, which is used by the subsequent OCCAM papers \citep[i.e,][]{occam_p4, myers_2022, otto_26}.
A Gaussian is then fit to the resulting distributions to determine individual membership probabilities for all proper motion members. 

The lower resolution data provided by the BOSS spectrographs presents challenges when determining bulk cluster parameters. 
Particularly in the case where there is only one or two stars in the cluster. 
In order for a cluster to appear in our final sample we require there to be at least 3 stars that have a greater than 14\% membership probability for all three membership criteria. 
This restriction avoids the case where there are 2 stars observed at plausible metallicities where it is nearly impossible to discern what the actual metallicity of the cluster is, and provides a higher level of confidence in the bulk parameters calculated with the low resolution data. 
We adopt an uncertainty floor in the bulk cluster abundance measurements equal to the standard deviation of the difference between the BOSS-CLAM and ASPCAP [Fe/H] values from DR19 for stars in both catalogs. 
In order to ensure reliable parameters we required all stars to have {\tt clam\_flags == 0} and {\tt snr} $>$ 30 in the BOSS-CLAM catalog and {\tt fe\_h\_flags == 0} and {\tt snr} $>$ 70 in the DR19 ASPCAP catalog. 

In addition to bulk cluster parameters (RV, [Fe/H]-abundance, [$\alpha$/M]-abundance) we also calculate several orbital parameters, including maximum distance from the plane, azimuth angle, eccentricity, orbital periods, and guiding center radius ($R_{Guide}$) using the {\tt gala} Galactic dynamics code \citep[][]{gala, gala_191}. 
\rguide is the radius of a circular orbit with the same angular momentum as the eccentric orbit, and has been shown to help account for orbital blurring effects in abundance gradients \citep[][]{Netopil21, spina_21, Zhang21}. 
We adopt \rguide as the primary radius for radial gradient analysis in this work. 

We combine the final sample of 111 clusters (95 new) with the DR19 OCCAM sample from \citet{otto_26} to constrain the radial Galactic [Fe/H] and [$\alpha$/M] gradients. 
Linear fits are performed using the {\tt emcee} package \citep[][]{emcee}, a Markov Chain Monte Carlo method software package. 
Values for the slope and y-intercept were estimated via maximum likelihood estimation, with uncertainties in each parameter estimated with the {\tt emcee} package.

\subsection{Changes from \citet{otto_26}}

The OCCAM survey made the switch from determining proper motion cluster members in-house to using the membership determinations based on \gaia 5-D astrometry \citep[e.g.,][]{cg20,HUNT_2023,Hunt2024} in \citet{otto_26}, which adopted the \citet{cg20} membership and cluster parameters.
In this work we use the cluster membership catalog from \citet{Hunt2024}. 
Eight clusters (NGC 2287, NGC 2516, NGC 2632, NGC 2682, NGC 3201, NGC 3532, NGC 4337, and Trumpler 10) were specifically targeted using the \citet{Hunt2024} member catalog which goes to lower magnitudes than \citet{cg20}, necessitating the switch to the newer catalog. Additionally, we use the cluster ages and distances from \citet{cavallo24}, who used the \citet{HUNT_2023} member catalog to re-derive cluster parameters with an artificial neural network. 

As this is the first OCCAM paper to use stellar parameters derived from BOSS spectra, it is the first to combine the current OCCAM cluster catalog with the previous one for analysis. 
In all plots where BOSS and APOGEE data appears, the BOSS DR20 clusters are shown as blue triangles and the APOGEE DR19 clusters are depicted as orange pentagons. 
The color scheme was adopted for visual clarity and because the previous color scheme (number of members in the cluster) was used as a visual quality metric, that no longer works as intended. 
Meaning, a cluster's parameters based on 10 BOSS stars are not necessarily of higher quality than cluster parameters derived from 5 APOGEE stars. 
We use representative error bars for each group of clusters.
The error on a cluster's radius is adopted as 5\% of the distance from the solar neighborhood to that cluster, while the y error is the median error of the cluster's abundance value ([Fe/H] or [$\alpha$/M]).

\section{OCCAM DR20 BOSS sample} \label{sec:boss_sample}

\begin{figure*}[ht]
  \epsscale{.9}
 \plotone{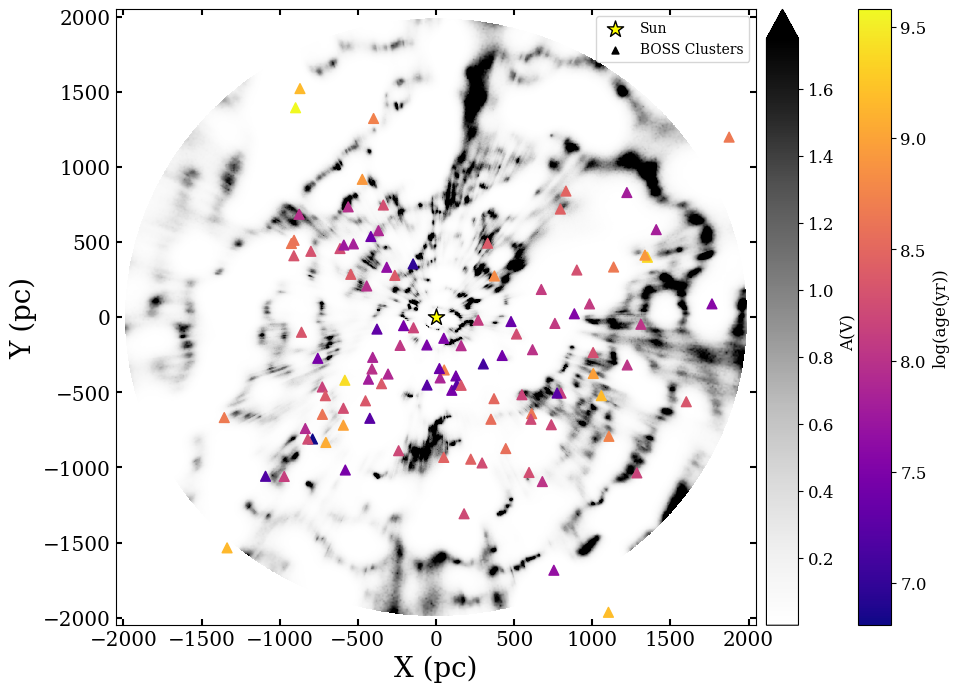}
 	\caption{ \small The BOSS OCCAM DR20 Solar neighborhood clusters plotted in the Galactic plane with the dust map from \citet{edenhofer2024}. Clusters are colored by log(age(yr)) and the dust map color bar saturates at a value of 1.75. The Sun is depicted as a yellow star at (0,0).} 
 	\label{fig:XYfull}
 \end{figure*}

The BOSS open cluster sample consists of 1883 stars in 111 different clusters, including 95 clusters that do not appear in the OCCAM DR19 sample from \citet{otto_26}, which primarily fills in the young cluster parameter space and are largely around the solar neighborhood.
We show the spatial distribution of the solar neighborhood BOSS clusters in Figure \ref{fig:XYfull}. 
The clusters are plotted on top of the extended 2 kpc dust map from \citet{edenhofer2024} and colored by the $log_{10}(age(yr))$ of each cluster. 
Since the BOSS OCCAM clusters are primarily located in the solar neighborhood, the ability to draw conclusions about large scale Galactic properties is limited while using only this sample.
For that reason, throughout this work, we combine the DR20 sample with the DR19 OCCAM sample from \citet{otto_26}, for a combined sample of 253 open clusters that span from 6-16 kpc in $R_{GC}$. 
Stars with membership probabilities greater than 14\% in all three membership criteria (proper motion\footnote{5D membership probability from \citet{Hunt2024}}, radial velocity and metallicity) were used to calculate the bulk BOSS DR20 cluster parameters reported in Table \ref{tab:full_params}, which is available as a machine-readable table.

\subsection{VAC information}

As in past OCCAM papers, two FITS tables have been made publicly available as a Value Added Catalog (VAC) in SDSS-V DR20.\footnote{The VAC files can be found at \url{data.sdss5.org/sas/dr20/vac/mwm/boss-occam/}} 
The first table, BOSS\_occam\_member-DR20-v1.fits, contains star IDs, positional information, stellar parameters (RV, $T_{eff}$, log(g)), chemistry, and membership probabilities for the stars with DR20 BOSS data in the \citet{Hunt2024} catalog for the 111 clusters in our final sample. 
This allows individual users to determine what membership threshold best suits their needs. 
The full list of columns included in the BOSS\_occam\_member file is shown in Table \ref{tab:vac}. 
The second table, BOSS\_occam\_cluster-DR20-v1.fits, reports bulk cluster parameters such as chemistry, motions, positions, and calculated orbital parameters for the 111 clusters. 
While new OCCAM VACs typically serve to replace previous versions, due to the lower resolution BOSS data, this VAC is best thought of as a supplement to the DR19 VAC, filling in the young cluster parameter space more fully. 
When a cluster appears in both catalogs, we recommend using the APOGEE based DR19 parameters.\footnote{With the exception of NGC 1901, which had insufficient data to correctly identify the cluster RV in DR19 resulting in the wrong star being chosen as the sole member of the cluster. With 6 member stars in DR20 this issue has been resolved.}

\begin{deluxetable}{ll}[ht!]
\tablecaption{A summary of the individual star data included in the DR20 OCCAM VAC \label{tab:vac}}
\tabletypesize{\scriptsize}
\tablehead{
    \colhead{Label} & 
    \colhead{Description}
    }
\startdata
Cluster & The associated open cluster \\
SDSS\_ID \tablenotemark{a}&   MWM {star} ID \\
GaiaDR3\_ID\tablenotemark{b} & \gaia DR3 star ID \\
GLON &  Galactic longitude \\
GLAT &  Galactic latitude \\
RAdeg &  right ascension \\
DEdeg & declination \\
V\_RAD \tablenotemark{c}& radial velocity \\ 
E\_V\_RAD\tablenotemark{c} & standard error in V\_RAD \\
PMRA\tablenotemark{b} &  proper motion in right ascension\\
E\_PMRA\tablenotemark{b}  & uncertainty in PMRA \\
PMDE\tablenotemark{b}  & proper motion in declination \\
E\_PMDE\tablenotemark{b}  & uncertainty in PMDEC \\
FeH\_CLAM\tablenotemark{d} & [Fe/H] from CLAM \\ 
E\_FeH\_CLAM\tablenotemark{d} & $1\sigma$ [Fe/H] dispersion \\
alpha\_M\_CLAM\tablenotemark{d} & [$\alpha$/M] from CLAM \\
E\_alpha\_M\_CLAM\tablenotemark{d} & $1\sigma$ [$\alpha$/M] dispersion \\
EH\_PROB &  membership probability from \\ & \citet{Hunt2024} \\
RV\_PROB &  membership probability based \\ &on RV (This study)\\
FEH\_PROB &  membership probability based \\ &on FE\_H\_CLAM (This study)\\
Teff\tablenotemark{d} & Effective temperature \\
logg\tablenotemark{d} & surface gravity log(g) \\[1ex]
\enddata
\tablenotetext{a}{Taken directly from MWM DR20.}\vskip-0.07in
\tablenotetext{b}{From \gaia DR3.} \vskip-0.07in
\tablenotetext{c}{From pyXCSAO \citet{pyxcsao}}
\tablenotetext{d}{From CLAM (I. Medan et al., {\it submitted})}
\end{deluxetable}

\noindent
\begin{deluxetable*}{lrrrrrrrrrrrrrcc}
\tabletypesize{\tiny}
\tablecaption{Basic Parameters and Chemistry of BOSS OCCAM DR20 Clusters \label{tab:full_params}}
\tablehead{
    \colhead{Cluster} &
    \colhead{l} &
    \colhead{b} &
    \colhead{RA} &
    \colhead{Dec} &
    \colhead{Radius\tablenotemark{a}} &
     \colhead{logAge\tablenotemark{b}} &
    \colhead{R$_{GC}$\tablenotemark{b}} & 
    \colhead{R$_{Guide}$\tablenotemark{c}} & 
    \colhead{$\mu_{\alpha}$\tablenotemark{a}} &
    \colhead{$\mu_{\delta}$\tablenotemark{a}} &
    \colhead{RV} & 
    \colhead{[Fe/H]} & 
    \colhead{[$\alpha$/M]} &
    \colhead{} & 
    \colhead{Num}\\[-4ex]
    \colhead{Name} &
    \colhead{(deg)} &
    \colhead{(deg)} &
    \colhead{(deg)} &
    \colhead{(deg)} &
    \colhead{(deg)} &
     \colhead{(dex)} &
    \colhead{(kpc)} &
    \colhead{(kpc)} &
    \colhead{(mas yr$^{-1}$)} &
    \colhead{(mas yr$^{-1}$)} & 
    \colhead{(km s$^{-1}$)} & 
    \colhead{(dex)} &
    \colhead{(dex)} &
    \colhead{Qual} &
    \colhead{Stars} 
    }
    \startdata
ASCC 127             &  112.47 &   +4.41 & 347.30 & +65.18 & 4.25 & 6.98 &  8.28 &  8.60 & $+7.41 \pm  0.02$ & $-1.65 \pm  0.02$ & $-13.3 \pm 4.90$ & $+0.21 \pm 0.02$ & $-0.02 \pm 0.02$ & 1 & 3\\
ASCC 58              &  281.70 &   +1.26 & 153.71 & -55.02 & 1.83 & 7.45 &  8.04 &  8.33 & $-13.30 \pm  0.02$ & $+2.76 \pm  0.02$ & $+10.6 \pm 3.80$ & $+0.00 \pm 0.02$ & $+0.03 \pm 0.03$ & 1 & 3\\
ASCC 71              &  299.99 &   -4.83 & 185.17 & -67.52 & 0.55 & 8.27 &  7.60 &  7.93 & $-9.36 \pm  0.01$ & $-1.36 \pm  0.01$ & $ -8.7 \pm 4.00$ & $-0.08 \pm 0.02$ & $+0.02 \pm 0.02$ & 1 & 4\\
Alessi-Teutsch 8     &  297.06 &   +1.36 & 180.70 & -60.95 & 0.52 & 8.57 &  7.72 &  8.55 & $-6.64 \pm  0.01$ & $+1.67 \pm  0.01$ & $-14.6 \pm 4.70$ & $-0.20 \pm 0.02$ & $+0.03 \pm 0.02$ & 1 & 3\\
Alessi 2             &  152.33 &   +6.34 &  71.56 & +55.18 & 1.52 & 8.40 &  8.67 &  9.44 & $-0.93 \pm  0.01$ & $-1.09 \pm  0.01$ & $-11.5 \pm 3.20$ & $-0.01 \pm 0.03$ & $+0.01 \pm 0.02$ & 2 & 9\\
Alessi 21            &  223.37 &   -0.08 & 107.59 &  -9.32 & 2.58 & 7.79 &  8.56 &  8.90 & $-5.50 \pm  0.01$ & $+2.65 \pm  0.01$ & $+39.5 \pm 4.70$ & $-0.07 \pm 0.03$ & $+0.05 \pm 0.04$ & 2 & 15\\
Alessi 24            &  328.93 &  -14.72 & 260.98 & -62.82 & 1.84 & 7.41 &  7.70 &  7.82 & $-0.43 \pm  0.01$ & $-8.96 \pm  0.01$ & $+11.2 \pm 4.10$ & $+0.09 \pm 0.04$ & $-0.03 \pm 0.02$ & 2 & 7\\
Alessi 31            &   15.36 &   +7.66 & 267.75 & -11.86 & 1.99 & 8.10 &  7.45 &  7.85 & $-1.13 \pm  0.01$ & $-4.17 \pm  0.01$ & $ +2.5 \pm 5.40$ & $-0.09 \pm 0.04$ & $+0.05 \pm 0.06$ & 2 & 5\\
Alessi 5             &  288.11 &   -1.97 & 160.90 & -61.11 & 1.56 & 7.41 &  8.00 &  8.18 & $-15.40 \pm  0.01$ & $+2.54 \pm  0.01$ & $+11.7 \pm 3.20$ & $-0.05 \pm 0.03$ & $+0.04 \pm 0.06$ & 2 & 7\\
Alessi 6             &  313.62 &   -5.54 & 220.01 & -66.14 & 1.00 & 8.62 &  7.54 &  7.51 & $-10.54 \pm  0.01$ & $-5.54 \pm  0.01$ & $-10.5 \pm 3.20$ & $-0.00 \pm 0.03$ & $-0.01 \pm 0.03$ & 2 & 7\\
\multicolumn{16}{c}{......}
    \enddata

\tablenotetext{a}{Bulk cluster parameters adopted from \citet{Hunt2024}}\vskip-0.07in
\tablenotetext{b}{Bulk cluster parameters adopted from \citet{cavallo24}}
\tablenotetext{c}{Calculated with distances from \citet{cavallo24}, computed with a solar radius of $R_{\odot} = 8.34$ kpc.}\vskip-0.07in
\tablenotetext{}{(This table is available in its entirety in machine-readable form.)}
\end{deluxetable*}

\section{Results} \label{results}

The DR20 BOSS data has added an additional 95 clusters, filling in the young cluster parameter space, that we use to supplement the DR19 APOGEE data to constrain the Galactic metallicity gradient\footnote{We also include the DR20 data instead of the DR19 data for NGC 1901, the cluster that has incorrect parameters in the DR19 catalog.}.  
We present the [Fe/H] and [$\alpha$/M] gradients using all 253 clusters, as well as gradients for different mono-age populations, in this section.
For a concise summary of all gradients reported in this section, see Table \ref{tab:gradients}. 

\subsection{Radial Trend in [Fe/H]}

\begin{figure*}[t!]
 	\begin{center}
         \epsscale{1.1}
     \plotone{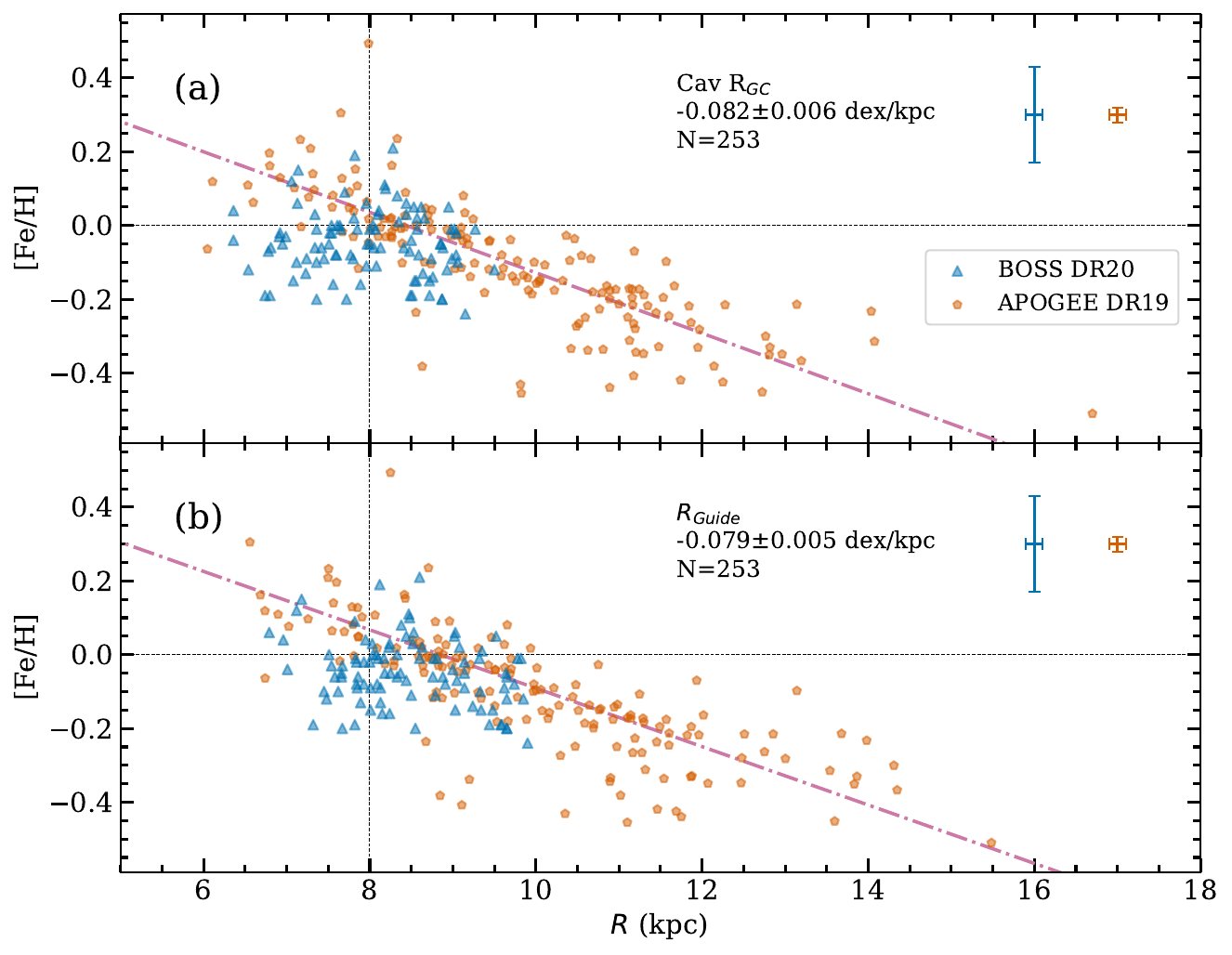} 
 	\end{center}
 	\caption{ \small The radial iron gradient with respect to both Galactocentric radius (top) and guiding center radius (bottom), using the combined cluster catalog, is represented by the pink dot-dashed line. Clusters with APOGEE DR19 measurements are shown as orange pentagons, and clusters with BOSS DR20 measurements are depicted as blue triangles. Representative error bars for each data set in the top right. }
 	\label{fig:feh_grad}
 \end{figure*}

The overall metallicity ([Fe/H]) gradients with respect to \rgc (a) and \rguide (b) are shown in Figure \ref{fig:feh_grad} with pink dot-dashed lines. 
\citet{otto_26} showed that the metallicity gradient, determined from the DR19 OCCAM sample, statistically preferred a linear fit instead of a bilinear fit with a knee. 
In this work, we only fit a linear gradient because we are supplementing the DR19 OCCAM sample with 95 new clusters, all of which fall inside the knee locations from \citet{otto_26}. 
We found the linear metallicity gradient with respect to \rgc to be $-0.082 \pm 0.006 \text{ dex kpc}^{-1}$. 
The metallicity gradient with respect to \rguide is slightly shallower at $-0.079 \pm 0.005 \text{ dex kpc}^{-1}$, but the two gradients are consistent with each other within their uncertainties. Both gradients are reported in Table \ref{tab:gradients}.

\begin{deluxetable}{llrrr}[h!]
\tabletypesize{\scriptsize}
\tablecaption{OCCAM DR20 Radial Gradients \label{tab:gradients}}
	\tablehead{
    \colhead{Abundance} &
    \colhead{Selection} &
    \colhead{Gradient} &
    \colhead{ N} &
    \colhead{Range}\\[-2.5ex]
    \colhead{Ratio} &
    \colhead{} &
    \colhead{(dex kpc$^{-1}$)} &
    \colhead{} &
    \colhead{(kpc)}\\[-4.5ex]
    }
\startdata
\multicolumn{5}{c}{\rgc}\\[0.5ex]\hline
{[Fe/H]}  &  All  & $-0.082 \pm 0.006$ & 253 & 6-18 \\
{[Fe/H]}  & $0.01 < \text{Age} \leq 0.15$ &   $-0.130 \pm 0.018$  & 63 & 6-13 \\
{[Fe/H]}  & $0.15 < \text{Age} \leq 0.4$  &   $-0.049 \pm 0.012$  & 62 & 6-13 \\
{[Fe/H]}  & $0.4 < \text{Age} \leq 0.8$   & $-0.089 \pm 0.011$   & 45 & 6-14 \\ 
{[Fe/H]}  & $0.8 < \text{Age} \leq 2.0$   & $-0.081 \pm 0.012$   & 45 & 6-13 \\
{[Fe/H]}  & $2.0 < \text{Age}$   & $-0.085 \pm 0.016$ & 29 & 6-18 \\\hline
{[$\alpha$/M]} & All & $+0.004 \pm 0.006$ & 253 & 6-18 \\
{[$\alpha$/M]}  & $0.01 < \text{Age} \leq 0.15$ &   $+0.003 \pm 0.021$  & 63 & 6-13 \\
{[$\alpha$/M]}  & $0.15 < \text{Age} \leq 0.4$  &   $-0.006 \pm 0.013$  & 62 & 6-13 \\
{[$\alpha$/M]}  & $0.4 < \text{Age} \leq 0.8$   & $+0.006 \pm 0.013$   & 45 & 6-14 \\ 
{[$\alpha$/M]}  & $0.8 < \text{Age} \leq 2.0$   & $+0.004 \pm 0.013$   & 45 & 6-13 \\
{[$\alpha$/M]}  & $2.0 < \text{Age}$   & $+0.011 \pm 0.017$ & 29 & 6-18 \\\hline
\multicolumn{5}{c}{\rguide}\\[0.5ex]\hline
{[Fe/H]}  &  All  & $-0.079 \pm 0.005$ & 253 & 6-16 \\
{[Fe/H]}  & $0.01 < \text{Age} \leq 0.15$ &   $-0.089 \pm 0.015$  & 63 & 6-13 \\
{[Fe/H]}  & $0.15 < \text{Age} \leq 0.4$  &   $-0.052 \pm 0.012$  & 62 & 6-15 \\
{[Fe/H]}  & $0.4 < \text{Age} \leq 0.8$   & $-0.080 \pm 0.011$   & 45 & 6-14 \\ 
{[Fe/H]}  & $0.8 < \text{Age} \leq 2.0$   & $-0.082 \pm 0.011$   & 45 & 6-14 \\
{[Fe/H]}  & $2.0 < \text{Age}$   & $-0.091 \pm 0.018$ & 29 & 6-16 \\\hline
{[$\alpha$/M]} & All & $+0.004 \pm 0.006$ & 253 & 6-16 \\
{[$\alpha$/M]}  & $0.01 < \text{Age} \leq 0.15$ &   $+0.011 \pm 0.018$  & 63 & 6-13 \\
{[$\alpha$/M]}  & $0.15 < \text{Age} \leq 0.4$  &   $-0.003 \pm 0.013$  & 62 & 6-15 \\
{[$\alpha$/M]}  & $0.4 < \text{Age} \leq 0.8$   & $+0.006 \pm 0.012$   & 45 & 6-14 \\ 
{[$\alpha$/M]}  & $0.8 < \text{Age} \leq 2.0$   & $+0.004 \pm 0.012$   & 45 & 6-14 \\
{[$\alpha$/M]}  & $2.0 < \text{Age}$   & $+0.011 \pm 0.018$ & 29 & 6-16 \\
\enddata

\end{deluxetable}

\subsection{Radial [$\alpha$/M] Trend}

BOSS-CLAM reports an $\alpha$-abundance ([$\alpha$/M]) in addition to [Fe/H]. 
While the DR19 OCCAM sample includes abundances for several $\alpha$ elements (O, Mg, Si, Ca), the [$\alpha$/M] values are derived using the ${\chi}^2$ minimization code {\tt FERRE} \citep[][]{AllendePrieto2006FERRE, aspcap} earlier in the ASPCAP pipeline are a better analog to the CLAM [$\alpha$/M]. 
We computed an average [$\alpha$/M] from the DR19 member stars in each cluster in order to combine the DR19 and DR20 OCCAM samples. 

We derive an identical radial [$\alpha$/M] gradient for both \rguide and \rgc at $+0.004 \pm 0.006 \text{ dex kpc}^{-1}$. 
The slightly positive gradient is consistent with no trend, meaning our results show that the $\alpha$-abundance decreases in step with overall metallicity as you move outward in the Galaxy.

\begin{figure*}
    \epsscale{1.1}
 	\plotone{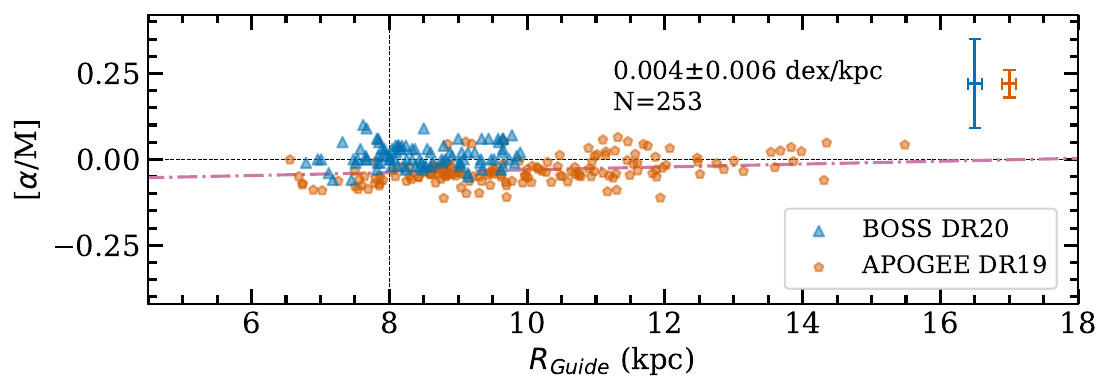}
 	\caption{ \small The [$\alpha$/M] vs \rguide radial gradient shown in the pink dot-dashed line. As in \ref{fig:feh_grad}, blue triangles are clusters with BOSS data and orange pentagons are clusters with APOGEE data. }
 	\label{fig:alpha}
 \end{figure*}

\subsection{Evolution of the Radial Gradients}
\subsubsection{Iron}

\begin{figure}
    \epsscale{1.2}
 	\plotone{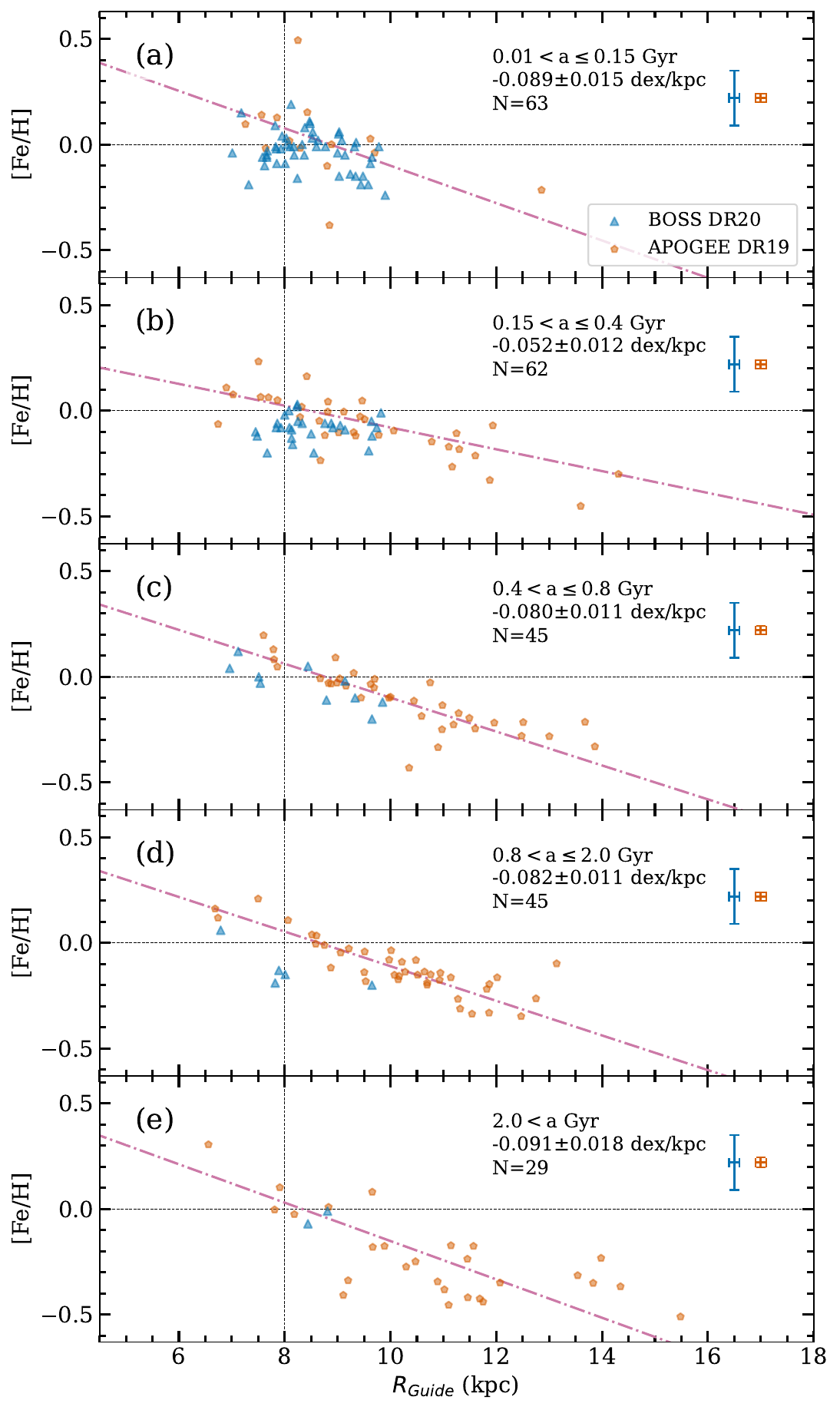}
 	\caption{ \small The radial Galactic [Fe/H] gradient for 5 age groups, (a) $0.01 \leq \text{age} \leq 0.15$ Gyr, (b) $0.15 \leq \text{age} \leq 0.4$ Gyr, (c) $0.4 \leq \text{age} \leq 0.8$ Gyr, (d) $0.8 \leq \text{age} \leq 2.0$ Gyr, and (e) age $> 2$ Gyr. Blue triangles are clusters with BOSS data, orange pentagons are clusters with APOGEE data and the overall gradient is shown with the pink dot-dashed line. }
 	\label{fig:age_grad}
 \end{figure}

In order to constrain the temporal evolution of the radial [Fe/H] gradient, we split our open cluster sample into five age bins (10 Myr $<$ age $\leq$ 150 Myr, 150 $<$ age $\leq$ 400 Myr, 400 $<$ age $\leq$ 800 Myr, 0.8 $<$ age $\leq$ 2.0 Gyr, age $>$ 2.0 Gyr), using the cluster ages from \citet{cavallo24}. 
Because the majority of the new DR20 clusters are younger than 500 Myr (92 out of 111) we split the youngest age bin from previous OCCAM papers into two age bins.
The radial [Fe/H] gradients with respect to \rguide are shown in Figure \ref{fig:age_grad}. 
The youngest age bin with 63 clusters spans radii from 6-13 kpc with a gradient of $-0.089 \pm 0.015 \text{ dex kpc}^{-1}$. 
The second youngest age bin has 62 clusters from 6-15 kpc with a gradient of $-0.052 \pm 0.012 \text{ dex kpc}^{-1}$, followed by a gradient of $-0.080 \pm 0.011 \text{ dex kpc}^{-1}$ for the 45 clusters in the middle age bin over a radius range of 6-14 kpc. 
The gradient for the second oldest age bin over the range of 6-14 kpc is $-0.082 \pm 0.011 \text{ dex kpc}^{-1}$ using 45 clusters. 
The oldest age bin has the steepest gradient, at $-0.091 \pm 0.017 \text{ dex kpc}^{-1}$, over the full range, 6-16 kpc, using 29 clusters. 
The gradients with respect to \rgc are recorded in Table \ref{tab:gradients} alongside those reported here. 

We see little evidence that the radial metallicity gradient with respect to \rguide varies with the age of the clusters it's derived from. 
Gradients in 4 out of 5 age bins are consistent with the overall linear gradient, with the exception of the second youngest age bin (150 $<$ age $\leq$ 400 Myr).

\subsubsection{[$\alpha$/M]}

\begin{figure}
    \epsscale{1.2}
 	\plotone{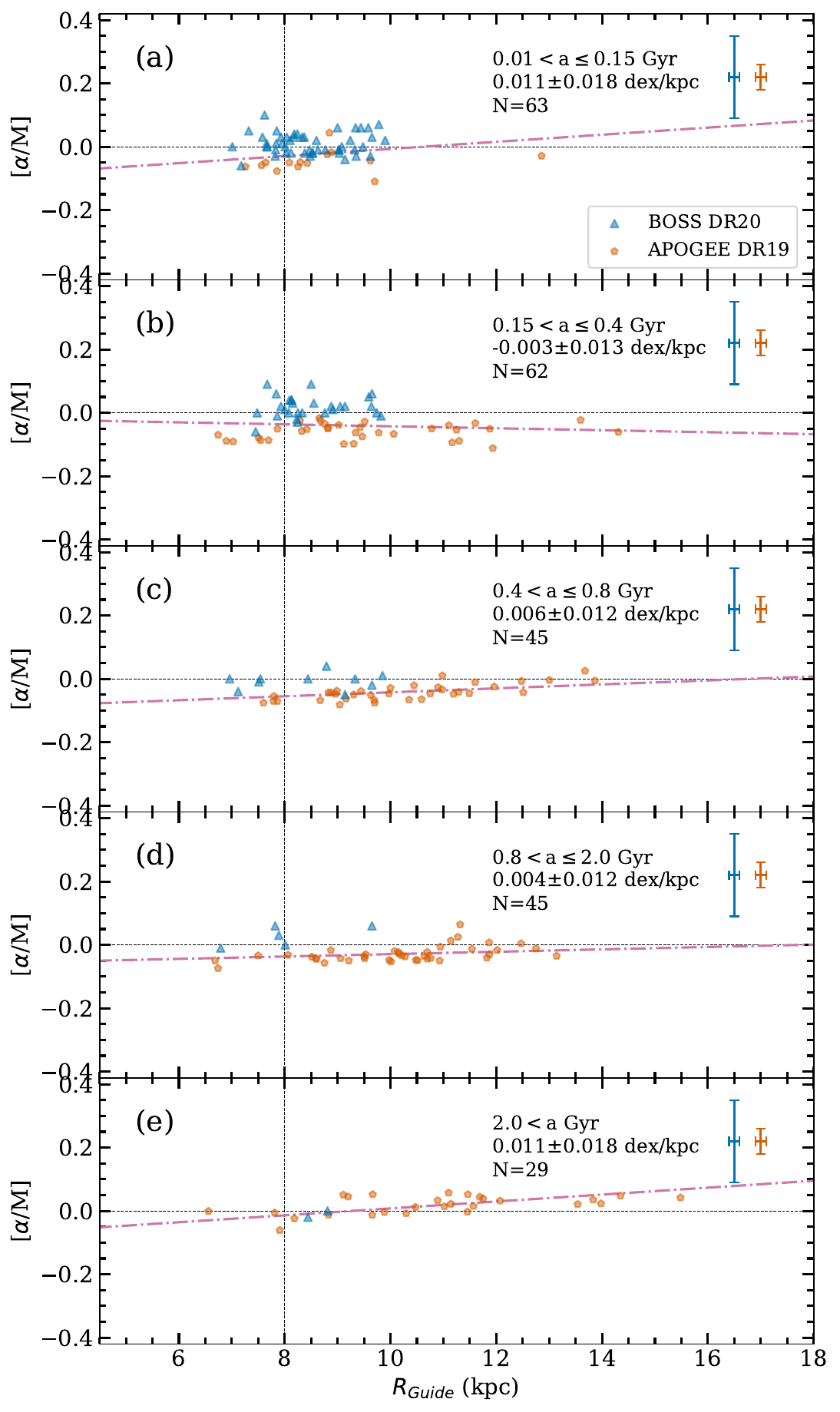}
 	\caption{ \small The radial Galactic [$\alpha$/M] gradient for 5 age groups, (a) $0.01 \leq \text{age} \leq 0.15$ Gyr, (b) $0.15 \leq \text{age} \leq 0.4$ Gyr, (c) $0.4 \leq \text{age} \leq 0.8$ Gyr, (d) $0.8 \leq \text{age} \leq 2.0$ Gyr, and (e) age $> 2$ Gyr. Blue triangles are clusters with BOSS data, orange pentagons are clusters with apogee data and the overall gradient is shown with the pink dot-dashed line. }
 	\label{fig:alpha_age_grad}
 \end{figure}

We use the same five age bins to investigate how the radial [$\alpha$/M] gradient with respect to \rguide has changed. 
Gradient values range from $-0.002$ to $+0.012 \text{ dex kpc}^{-1}$ but all age bins have gradients consistent with no trend as seen in Figure \ref{fig:alpha_age_grad}. 
Like the overall [$\alpha$/M] gradient the age bin $\alpha$ gradients follow the variations in the [Fe/H] gradients within each age bin, even though the [Fe/H] gradient diverges significantly from the overall [Fe/H] trend in specific age bins. 
The gradients with respect to \rguide and \rgc are recorded in Table \ref{tab:gradients}.

\section{Discussion} \label{sec:discussion}

\subsection{Comparison to \citet{otto_26}}

The radial [Fe/H] gradients with respect to both \rguide ($-0.079 \pm 0.005 \text{ dex kpc}^{-1}$) and \rgc ($-0.082 \pm 0.006 \text{ dex kpc}^{-1}$) agree superbly with their counterparts from \citet{otto_26} ($-0.071 \pm 0.006 \text{ dex kpc}^{-1}$ and $-0.079 \pm 0.006 \text{ dex kpc}^{-1}$) which only used the OCCAM DR19 sample. 
The addition of 95 new clusters, 86 with ages less than 500 Myr, does not significantly alter the radial metallicity gradient inferred from the larger sample.
While this behavior is not wholly unexpected due to the APOGEE clusters with smaller errors having a greater influence on the fitting routine than the BOSS clusters with larger errors, this is still an important result.

\subsection{Comparison to other Surveys}

We also compare our gradients to those found in \citet{spina_21} as a high-resolution optical control and \citet{yang2025} as a low-resolution optical analog. 
\citet{spina_21} used a sample of 134 open clusters that range from 6-16 kpc in \rguide. While \citet{yang2025} had a slightly larger sample of 204 clusters with a smaller range in \rgc, from 6-12 kpc.  

\subsubsection{[Fe/H]}

\citet{spina_21} report an [Fe/H] gradient with respect to \rguide of $-0.073 \pm 0.008 \text{ dex kpc}^{-1}$ which is in good agreement with the value we find in this work, $-0.079 \pm 0.005 \text{ dex kpc}^{-1}$. 
Alternatively, \citet{yang2025} report a shallower slope with respect to \rgc of $-0.048 \pm 0.008 \text{ dex kpc}^{-1}$. 
When we restrict our sample of clusters to the same range that they use (6-12 kpc), we find an [Fe/H] gradient of $-0.083 \pm 0.006 \text{ dex kpc}^{-1}$, nearly identical to the value when using the whole sample. 
It is also highly discrepant with the value from \citet{yang2025} but agrees well with the slope with respect to \rgc from \citet{spina_21}, $-0.076 \pm 0.009 \text{ dex kpc}^{-1}$. 

As both \citet{yang2025} and this work used abundances derived from low resolution spectra using label transfer routines  (e.g., the Payne \citep[][]{Ting2019} for \citet{yang2025} and BOSS-CLAM here) we would expect the overall gradient to be in better agreement. 
Different clusters in each sample could provide an explanation to this discrepancy as fitting lines to different data points could result in different slopes being found as there are only 13 clusters in common. 
But it is also possible that the BOSS-CLAM methodology offers real improvement over previous label transfer methods such as the Payne. 
Currently, there are only 13 clusters in common between the OCCAM BOSS sample and the LAMOST sample of open clusters, which do show good agreement with a median offset of $0.04 \pm 0.08\text{ dex kpc}^{-1}$. 
As the BOSS OCCAM sample continues to grow in future SDSS-V data releases we hope to provide clarity to whether abundance determination methodology, sample selection, or another factor is the driving force behind the discrepant gradients.

\subsubsection{[$\alpha$/M]}

Our [$\alpha$/M] gradient with respect to \rguide is consistent with no trend which agrees well with three of the [$\alpha$/Fe] gradients available in \citet{spina_21} (Mg, Si, and Ca). 
O and Ti both exhibit positive gradients in \citet{spina_21}. 
In future analyses of BOSS spectra, we hope to gain individual $\alpha$-abundance values which will allow us to investigate whether we can recover the trends seen in individual elements by the higher resolution survey.

\subsection{Comparisons to other Tracers of the Galactic Radial Gradient}

We leverage the large increase in clusters younger than 500 Myr to compare the radial metallicity gradient as seen with open clusters to other tracers of the Galactic metallicity gradient, namely cepheid stars and HII regions, in this section. 

\subsubsection{Cepheid Stars}

\begin{figure}[htb]
    \epsscale{1.2}
 	\plotone{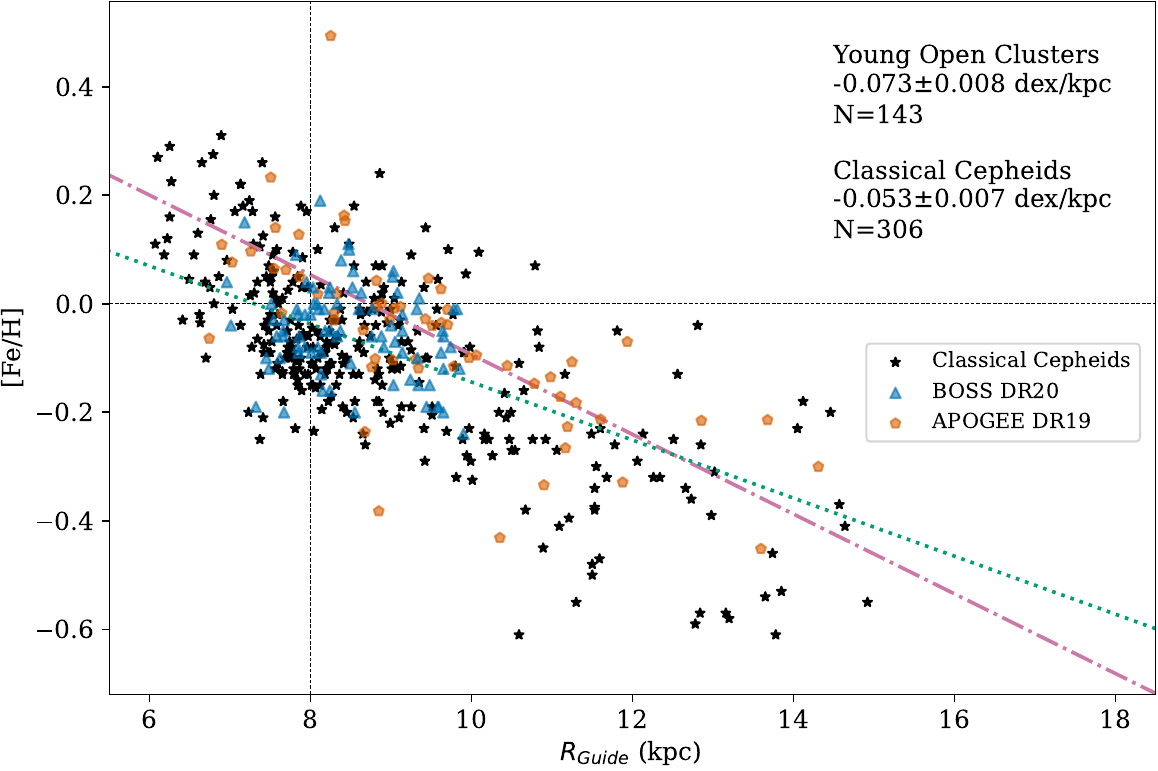}
 	\caption{ \small The CCs (black stars) from \citet{nunnari_26} and young open clusters (blue triangles and orange pentagons) over-plotted with their respective radial gradients. 
    The pink dot-dashed line shows the gradient from the young open-clusters, while the green dotted line is the gradient calculated from the CCs. }
 	\label{fig:ccs_ocs}
 \end{figure}

Classical Cepheids (CCs) make excellent tracers of young stellar populations as they are intermediate-mass stars that are less than a few hundred million years old with well constrained distances \citep[][]{bono_2024}. 
We compare clusters younger than 500 Myr, excluding any clusters with \citet{cavallo24} ages less than 10 Myr, to the sample of 401 CCs from \citet{nunnari_26}.
In order to make a better comparison, we compute the guiding center radius for the CCs in the \citet{nunnari_26} sample, and restrict the radius range to 6-15 kpc to correspond to the open cluster sample. 
We report a radial metallicity gradient of $-0.053 \pm 0.007 \text{ dex kpc}^{-1}$ for the 306 CCs that meet this criteria. 
The 143 young open clusters indicate a steeper radial gradient at $-0.073 \pm 0.008 \text{ dex kpc}^{-1}$. 
Both populations and their gradients are shown in Figure \ref{fig:ccs_ocs} for a visual comparison. 
The reason for this discrepancy is not immediately apparent but a likely contributing factor is the distribution of the two samples in \rguide space. 
Only 18 open clusters ($\sim13\%$) have guiding center radii larger than 10 kpc, while 66 CCs ($\sim22\%$) are in that same region. 
If there is a break in the radial metallicity gradient past 10 kpc the current sample of young open clusters will not probe that as well as the sample of CCs. 
Potentially resulting in the steeper slope we see in the data. 

\subsubsection{H II Regions}

H II regions are an excellent probe of the current abundance gradient of the interstellar medium (ISM) \citep[][]{martin_2017}. 
\citet{mendezdelgado_22} used a sample of 42 H II regions from \citet{arellano_20, arellano_21} and \citet{md_20} to derive the radial abundance gradient for several elements, including [O/H], a proxy for total metallicity. 
\citet{mendezdelgado_22} report two different abundances for their H II regions one determined with the standard direct method ($t^2 = 0$) and one where they account for the internal temperature inhomogeneities ($t^2 > 0$). 
We make use of the $t^2 > 0$ abundances for comparison in this work. 
We derive the radial metallicity gradient using the H II regions in that sample that fall within the \rgc range where we have open clusters (6-14 kpc)\footnote{We use \rgc for the H II regions instead of \rguide because of the lack of widely available, reliable radial velocity measurements. Which are necessary for computing $\text{R}_{Guide}$.}, finding a radial gradient of $-0.053 \pm 0.015 \text{ dex kpc}^{-1}$. 
This gradient is shallower than the one we see with the 143 young open clusters, $-0.081 \pm 0.009 \text{ dex kpc}^{-1}$, but is nearly identical to the gradient with respect to \rgc as determined by the CCS, $-0.053 \pm 0.007 \text{ dex kpc}^{-1}$. 
Having the two young tracers agree so well while the young open clusters disagree calls into question whether young open clusters trace the same ISM material that H II regions and CCs do. 
Which could be the case if the cluster ages are incorrect.

\subsection{Time Evolution}

By splitting our sample of clusters into different mono-age populations we can constrain how the radial metallicity gradient has changed over time. 
We find that when using our sample of open clusters, the radial metallicity gradient is nearly constant across all 5 mono-age populations. 
Only the population with clusters between 150 Myr and 400 Myr has a radial gradient that disagrees with the overall radial gradient. 
Our results lend support to the equilibrium scenario proposed in \citet{johnson_25}, where the interstellar medium (ISM) metallicity, quickly evolves to a local equilibrium abundance at fixed Galactic radii. 
Which will result in a radial metallicity gradient that does not change significantly within different mono-age populations. 
We note that even though we have substantially increased the number of open clusters in the sample our results mirror the results from \citet{otto_26}, where 3 out of 4 mono-age bins had gradients consistent with the overall linear gradient using only the APOGEE clusters. 
This is especially striking as 86 of the 95 added clusters have ages less than 500 Myr but the youngest age bin still indicates little to no change in the radial metallicity gradient. 

However, there are several caveats that complicate the interpretation of our results. 
Young stars, particularly member stars of the clusters in the youngest age bin ($10 < \text{ age } < 150$ Myr) can be chromospherically active with emission lines that complicate metallicity determinations \citep[][]{ystar_activity_09}. 
Such young stars are flagged in the BOSS-CLAM as potentially unreliable ({\tt flag\_warn = False}) if they were targeted as part of the {\tt mwm\_ob} program, or bad ({\tt flag\_bad = True}) if they were targeted in the {\tt mwm\_yso} program, to mitigate these issues.
Given the large number of stars in this age group, it is likely that some subset have unreliable [Fe/H] values. 
Alternatively, if we trust the {\tt clam\_flags} it calls into question the veracity of the cluster ages.  

\citet{cavallo24} allowed the metallicity of the cluster to be a free parameter during their isochrone fitting routine. 
When compared with the spectroscopic [Fe/H] abundances for the APOGEE OCCAM DR19 sample of 158 clusters we find that 124 have differences larger than 0.1 dex, 80 have differences larger than 0.25 dex, and three clusters have a [Fe/H] discrepancy of over 1.0 dex. 
While \citet{cavallo24} show that their ages agree well with the ages from \citet{cg20}, such a large discrepancy between the metallicity of the cluster can alter the isochrone-derived ages.

Whether the oldest population paint an accurate picture of the radial gradient in the disk 2 Gyr ago is up for debate. 
Old open clusters ($> 3$ Gyr) have been shown to have more perturbed orbits than their younger counterparts \citep[][]{vv_radmig_23}. 
They are also more rare than their younger counterparts. 
The consequence of this being that it is likely that old open clusters do not accurately trace the radial gradient of their present day location at the time of their birth \citep[][]{vv_radmig_23}. 
Using \rguide instead of \rgc can help mitigate some of the radial migration effects, but as \citet{bob_25} showed old open clusters in FIRE simulated galaxies can undergo significant changes in their angular momentum, which will affect the calculated guiding center radius.

\section{Conclusions}
\label{sec:conclusions}

We have produced a catalog of 111 clusters from SDSS-V/MWM DR20 from stars with low-resolution optical data. 
This catalog serves as a supplement to the DR19 OCCAM sample from \citet{otto_26}, and adds 95 new clusters, 86 of which are younger than 500 Myr. 
We use the catalog to constrain the radial Galactic metallicity gradient and find a gradient with respect to \rguide of $-0.079 \pm 0.005 \text{ dex kpc}^{-1}$ and a gradient with respect to \rgc of $-0.082 \pm 0.006 \text{ dex kpc}^{-1}$. 
The addition of the predominantly young BOSS clusters does not significantly change the radial gradient of the Milky Way as compared to the older APOGEE only sample. 
We find that our gradient in \rguide agrees well with the gradient from \citet{spina_21} using 134 clusters from GALAH and APOGEE, but we find a steeper slope than \citet{yang2025} does using data from LAMOST. 
We leverage the large sample of young open clusters (143) with ages between 10 Myr and 500 Myr to directly compare to the metallicity gradients obtained from H II regions and classical cepheid stars. 
The young open clusters indicate a steeper gradient in \rgc space ($-0.081 \pm 0.009 \text{ dex kpc}^{-1}$) than either the classical cepheids ($-0.053 \pm 0.007 \text{ dex kpc}^{-1}$) or the H II regions ($-0.053 \pm 0.015 \text{ dex kpc}^{-1}$). 
When determining the radial metallicity gradient for 5 mono-age populations, we find that the gradient stays mostly constant, favoring the equilibrium scenario described in \citet{johnson_25}. 
This was also seen using the APOGEE only sample from \citet{otto_26}. 
We again highlight how adding a large sample of clusters, 95 total, 86 of which are younger than 500 Myr, has not changed the result that our open cluster sample points to a static Galactic radial gradient.  
Larger samples, particularly at larger Galactic radii and older ages are needed to assess the veracity of these results and further investigate how radial migration plays a role in determining the inferred Galactic metallicity gradient.

\begin{acknowledgments}
JMO, PMF, JD, AH, NRM, KC, AS, and GZ acknowledge support for this research from the National Science Foundation Astronomy and Astrophysics grant AST-2206541, AST-2206542, and AST-2206543.

J.E.M.-D. gratefully acknowledges support from the Secretaria de Ciencia, Humanidades, Tecnologia e Innovacion (SECIHTI) project CBF-2025-I-2048, ``Resolviendo la Fisica Interna de las Galaxias: De las Escalas Locales a la Estructura Global con el SDSS-V Local Volume Mapper", and from the UNAM/DGAPA/PAPIIT project IA103326, ``DESIRED (DEep Spectra of ionised Regions Database): de las emisiones mas sutiles a la fisica fundamental del universo".

J.G.F-T gratefully acknowledges the support provided by ANID Fondecyt Regular No. 1260371

Funding for the Sloan Digital Sky Survey V has been provided by the Alfred P. Sloan Foundation, the Heising-Simons Foundation, the National Science Foundation, and the Participating Institutions. SDSS acknowledges support and resources from the Center for High-Performance Computing at the University of Utah. SDSS telescopes are located at Apache Point Observatory, funded by the Astrophysical Research Consortium and operated by New Mexico State University, and at Las Campanas Observatory, operated by the Carnegie Institution for Science. The SDSS website is \url{www.sdss.org}.

SDSS is managed by the Astrophysical Research Consortium for the Participating Institutions of the SDSS Collaboration, including Caltech, The Carnegie Institution for Science, Chilean National Time Allocation Committee (CNTAC) ratified researchers, The Flatiron Institute, the Gotham Participation Group, Harvard University, Heidelberg University, The Johns Hopkins University, L’Ecole polytechnique f\'{e}d\'{e}rale de Lausanne (EPFL), Leibniz-Institut f\"{u}r Astrophysik Potsdam (AIP), Max-Planck-Institut f\"{u}r Astronomie (MPIA Heidelberg), Max-Planck-Institut f\"{u}r Extraterrestrische Physik (MPE), Nanjing University, National Astronomical Observatories of China (NAOC), New Mexico State University, The Ohio State University, Pennsylvania State University, Smithsonian Astrophysical Observatory, Space Telescope Science Institute (STScI), the Stellar Astrophysics Participation Group, Universidad Nacional Aut\'{o}noma de M\'{e}xico, University of Arizona, University of Colorado Boulder, University of Illinois at Urbana-Champaign, University of Toronto, University of Utah, University of Virginia, Yale University, and Yunnan University.\end{acknowledgments}

\begin{contribution}

JMO led the DR20 cluster analysis and is the primary author of the paper. 
PMF leads the OCCAM survey and contributed to the data analysis, interpretation of results and writing of the paper. 
JD contributed to the development of the OCCAM pipeline and writing of the paper. 
IM led the production of the BOSS-CLAM abundance catalog and contributed to the writing of the paper. 
NRM and AH contributed to the vetting of the cluster sample and writing of the paper. 
KC provided expertise in evaluating the quality of the abundances and comments on the draft. 
GZ and MM contributed discussions that shaped the scope and direction of the project. 
JEM-D and JZ contributed to the comparisons of other tracers and provided comments on the draft. 
AS, ZW and JGF-T contributed to discussions about the paper and provided comments on the draft.
SDSS-V Architects (Dmitry Bizyaev, Andrew R. Casey) contributed more than 1 FTE to the fund-raising, proposal writing, hardware, software, engineering, operations, observing, data archiving and/or scientific organization of SDSS-V from which the data from this work are derived. 

\end{contribution}

\facilities{SDSS}

\software{{\tt astropy} \citep{2013A&A...558A..33A,2018AJ....156..123A, astropy:2022}, {\tt dustmaps} \citep[][]{dustmaps} , {\tt emcee} \citep[][]{emcee} , {\tt gala} \citep[][]{gala,gala_191}, {\tt matplotlib} \citep{matplotlib}, {\tt numpy} \citep{numpy}, {\tt pandas} \citep[][]{pandas_cf, pandasv233}, {\tt scipy} \citep[specifically {\tt curve\_fit},][]{scipy, curve_fit}}

\appendix

\section{Post Hoc [Fe/H] Calibrations}\label{app:cal}

\begin{figure*}[htb]
    \centering
    \epsscale{1.0}
    \includegraphics[width=.95\textwidth]{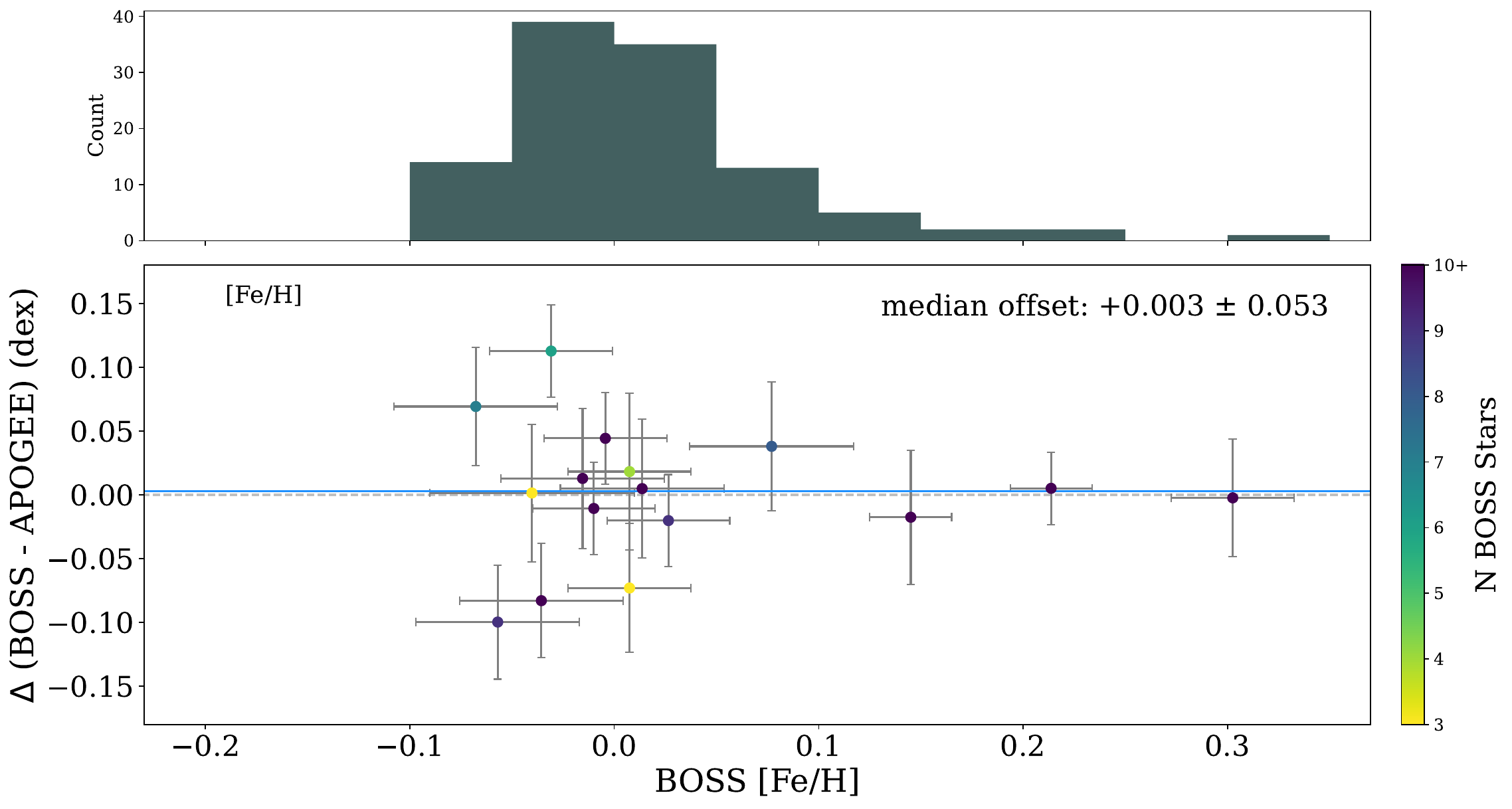}
    \caption{\small The DR20 BOSS minus DR19 APOGEE delta plot for [Fe/H] using the corrected DR20 [Fe/H] abundances for the 16 overlap clusters. 
    Points are colored by the number of stars in the cluster in the DR20 OCCAM catalog.
    The grey dashed line shows the zero point and the solid blue line shows the median offset. 
    The [Fe/H] distribution of the entire corrected DR20 OCCAM sample is shown above the scatter plot.}
    \label{fig:cor_comp}
\end{figure*}

In an effort to correct the trend seen in Figure \ref{fig:comp}, where the difference between the BOSS and APOGEE bulk cluster [Fe/H] abundances becomes increasingly negative as a function of metallicity, we fit a quadratic function to the data points from the top scatter plot of Figure \ref{fig:comp}. 
The resulting equation:
\begin{equation}
    y = -0.87x^{2} + 0.32x -0.03
\end{equation}
was used to first correct the [Fe/H] abundances of the 16 BOSS clusters in the overlap sample and then applied to the whole sample to test if the correction changed the inferred gradients. 
Figure \ref{fig:cor_comp} shows the delta plot for the overlap sample with the corrected DR20 [Fe/H] abundances. 
The median offset did decrease to nearly zero (+0.003) largely due to the clump of seven clusters at an [Fe/H] of $\sim-0.5$ dex in Figure \ref{fig:comp} that the fit passed through, being brought up to an offset of zero.
However, this results in a bifurcation at the low metallicity end where several clusters diverge to lower [Fe/H] values than their APOGEE counterparts while several others go to higher [Fe/H] values. 
We also refit both the overall metallicity gradient in both \rguide and \rgc and the gradient with respect to \rguide in our 5 mono-age population bins (shown in Figures )and found that the corrections did not significantly impact the inferred gradients. 
Because this correction did not affect our inferred gradients nor the conclusions drawn from them and seems to introduce new systematics, while being based on only 16 clusters, we opted not to use the correction in this work. 
However, we do note that when analyzing a single cluster, or a small subset of clusters, a careful treatment of the systematics seen in Figure \ref{fig:comp} is likely necessary.

\begin{figure*}[htb]
 	\begin{center}
         \epsscale{1.1}
     \plotone{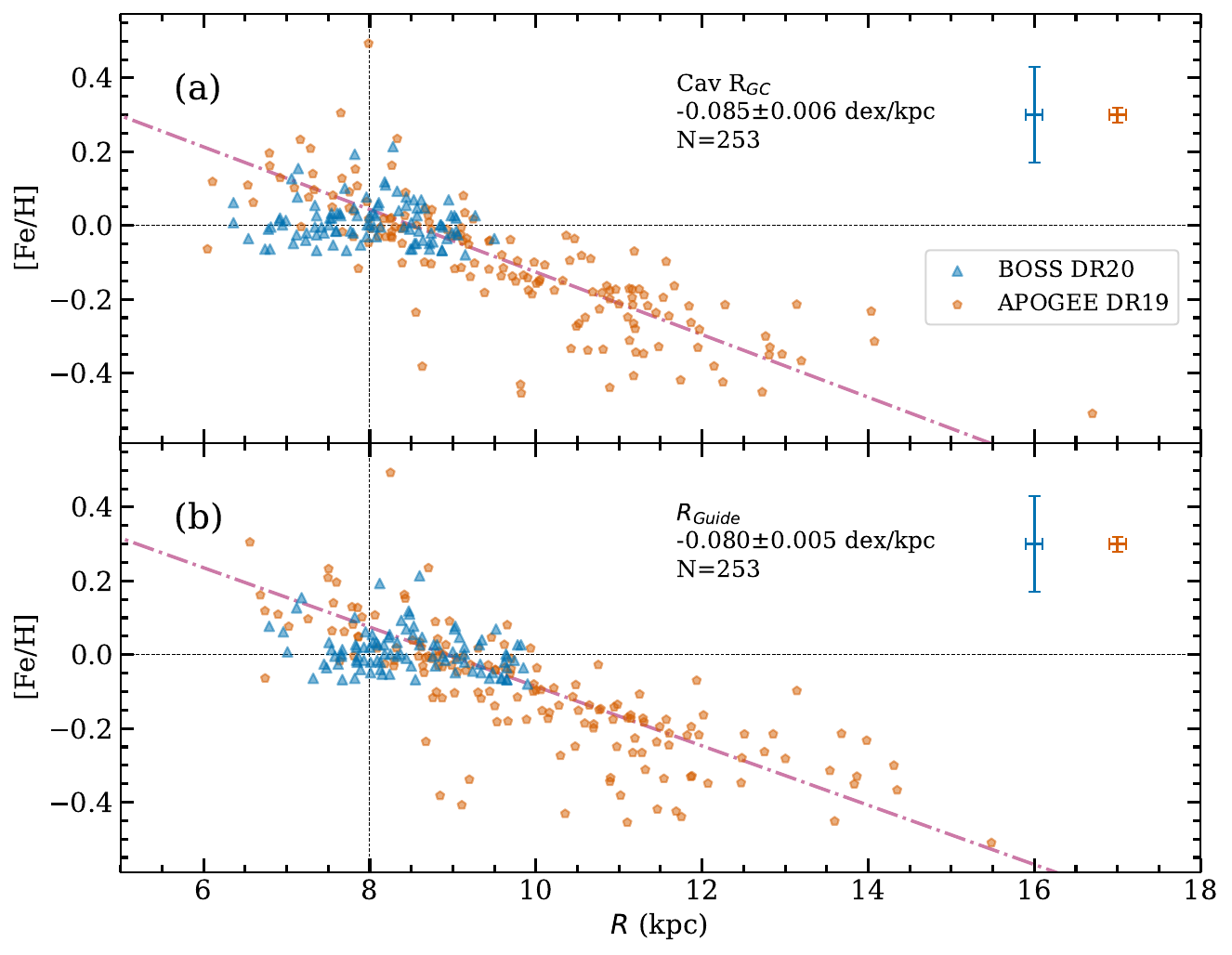} 
 	\end{center}
 	\caption{ \small Same as Figure \ref{fig:feh_grad} but with the calibrated DR20 [Fe/H] abundances. }
 	\label{fig:feh_grad_calib}
 \end{figure*}

 \begin{figure}[htb]
    \epsscale{1.2}
 	\plotone{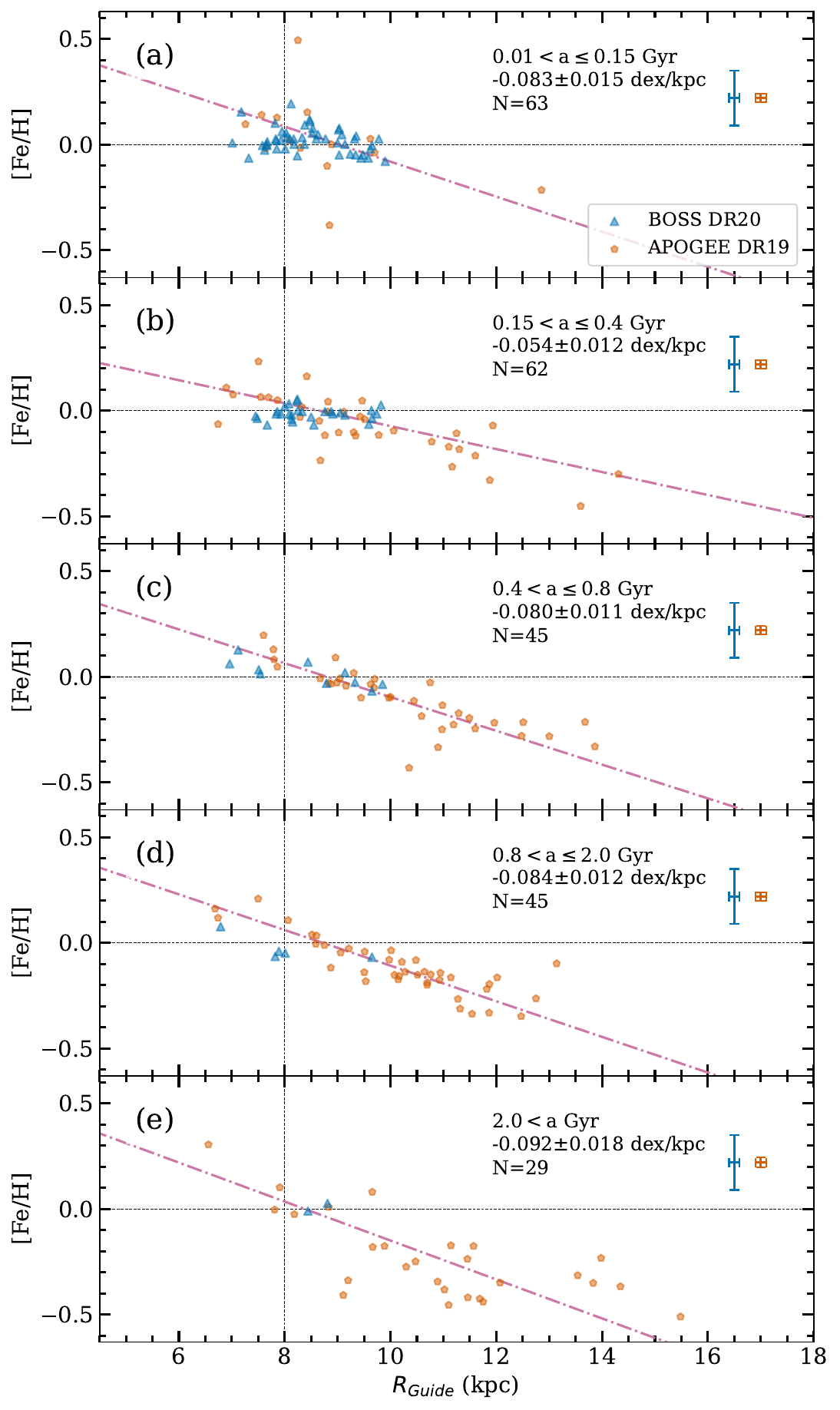}
 	\caption{ \small Same as Figure \ref{fig:age_grad} but with the calibrated DR20 [Fe/H] abundances. }
 	\label{fig:age_grad_calib}
 \end{figure}

\bibliography{dr20}{}
\bibliographystyle{aasjournalv7}

\end{document}